\documentclass{aastex}                 
\usepackage{emulateapj5,apjfonts,epsf}
 
\received{}
\accepted{}
\journalid{}{}
\articleid{}{}

\journalinfo{{\sc The Astrophysical Journal}, 2002, in press}

\slugcomment{Accepted for publication in The Astrophysical Journal
Letters April 12, 2002}

\shortauthors{Camilo et al.}
\shorttitle{Radio pulsar in SNR~3C58}

\begin{document}

%
%

\def\psr{PSR~J0205+6449}
\def\chandra{{\em Chandra\/}}

\title{Discovery of radio pulsations from the X-ray pulsar J0205+6449
in supernova remnant 3C58 with the Green Bank Telescope }

\author{F.~Camilo,\altaffilmark{1} 
  I.~H.~Stairs,\altaffilmark{2}
  D.~R.~Lorimer,\altaffilmark{3}
  D.~C.~Backer,\altaffilmark{4}
  S.~M.~Ransom,\altaffilmark{5, 6}
  B.~Klein,\altaffilmark{7}
  R.~Wielebinski,\altaffilmark{7}
  M.~Kramer,\altaffilmark{3}
  M.~A.~McLaughlin,\altaffilmark{3}
  Z.~Arzoumanian,\altaffilmark{8} and
  P.~M\"uller\altaffilmark{7} }
\altaffiltext{1}{Columbia Astrophysics Laboratory, Columbia University,
  550 West 120th Street, New York, NY~10027}
\altaffiltext{2}{National Radio Astronomy Observatory, P.O. Box 2,
  Green Bank, WV~24944}
\altaffiltext{3}{University of Manchester, Jodrell Bank Observatory,
  Macclesfield, Cheshire, SK11~9DL, UK}
\altaffiltext{4}{Astronomy Department, University of California,
  Berkeley, CA~94720}
\altaffiltext{5}{Physics Department, McGill University, 3600 University
  Street, Montreal, QC, H3A~2T8, Canada}
\altaffiltext{6}{Center for Space Research, Massachusetts Institute of
  Technology, Cambridge, MA~02139}
\altaffiltext{7}{Max-Planck-Institut f\"ur Radioastronomie, Auf dem
  H\"ugel 69, D-53121, Bonn, Germany}
\altaffiltext{8}{Universities Space Research Association/NASA-GSFC,
  Code 662, Greenbelt, MD~20771}

\begin{abstract}
We report the discovery with the 100\,m Green Bank Telescope of 65\,ms
radio pulsations from the X-ray pulsar J0205+6449 at the center of
supernova remnant 3C58, making this possibly the youngest radio pulsar
known.  From our observations at frequencies of 820 and 1375\,MHz, the
free electron column density to \psr\ is found to be $140.7 \pm
0.3$\,cm$^{-3}$\,pc.  The barycentric pulsar period $P$ and $\dot P$
determined from a phase-coherent timing solution are consistent with
the values previously measured from X-ray observations.  The averaged
radio profile of \psr\ consists of one sharp pulse of width $\approx
3\,\mbox{ms} \approx 0.05 P$.  The pulsar is an exceedingly weak radio
source, with pulse-averaged flux density in the 1400\,MHz band of
$\sim 45\,\mu$Jy and a spectral index of $\sim -2.1$.  Its radio
luminosity of $\sim 0.5$\,mJy\,kpc$^2$ at 1400\,MHz is lower than that
of $\sim 99\%$ of known pulsars and is the lowest among known young
pulsars.

\end{abstract}

\keywords{ISM: individual (3C58) --- pulsars: individual (\psr) ---
supernova remnants}

\section{Introduction}\label{sec:intro} 

The ``Crab-like'' supernova remnant (SNR) G130.7+3.1 (3C58) has long
been suspected of containing a pulsar at its center (e.g., Becker,
Helfand, \& Szymkowiak 1982\nocite{bhs82}).  This SNR is of particular
interest because it is thought to be the remnant of the explosion
recorded in 1181~AD (SN~1181), and hence to have a known age of
820\,yr.  After more than 20 years of searching in the X-ray and radio
bands, the pulsar J0205+6449 with period $P=65.68$\,ms was recently
discovered at the SNR center with the \chandra\ X-ray Observatory
\cite{mss+01}.

The historical record associating the pulsar-powered synchrotron nebula
3C58 with SN~1181 is compelling.  Nevertheless, 3C58 has some
properties suggesting a larger age (see, e.g., Slane, Helfand, \&
Murray 2002\nocite{shm02}, and references therein).  Also, the nebular
radio emission of 3C58 is $\sim 10 \times$ lower than the comparably
aged Crab Nebula's, while the X-ray emission is $\sim 1000 \times$
weaker.  Because the synchrotron lifetime of X-ray-emitting electrons
is short (unlike of those generating radio emission), it is surprising
that the ratio of the X-ray luminosities from the 2 SNRs does not match
the relative present-day spin-down luminosities of the
pulsars:  $\dot E_{\rm Crab}/\dot E_{\rm J0205+6449} \approx 15$.
Since $\dot E \propto \dot P/P^3$, the power injected into the nebula
as a function of time depends strongly on $P(t)$, which therefore is an
important quantity to constrain; the subsequent nebular evolution is
governed in addition by the pressure of the ambient medium.  The
characteristic age $\tau_c = P/2\dot P = 5400$\,yr of J0205+6449 can be
reconciled with the 1181~AD explosion by appealing to a large initial
spin period of $\sim 60$\,ms \cite{mss+01}.  However we do not know the
form of the spin evolution of the pulsar: measuring its ``braking
index'' $n$, where the angular frequency of rotation evolves as $\dot
\Omega \propto - \Omega^n$, would constrain $P(t)$ and help to
understand the energetics of 3C58.

The study of J0205+6449 at radio wavelengths could contribute
significantly towards the determination of the braking index: precise
long-term monitoring of the pulsar rotation, especially important for
young pulsars which often experience period glitches, may prove easier
from the ground than from satellites.  In addition, the absolute phase
alignment of radio and X-ray pulses should provide information
regarding the emission mechanism.  A radio detection would also yield
the dispersion measure (DM), giving an independent distance constraint
or, alternatively, the average free electron density along the line of
sight.  Finally, measuring the radio luminosity and spectrum of very
young pulsars is essential for constraining the population of these
objects in the Galaxy.  Motivated by these factors we made deep
searches for radio pulsations from \psr, culminating in their discovery
with the new Green Bank Telescope as reported here.

\section{Observations}\label{sec:obs}

The SNR~3C58 was most recently searched for a radio pulsar by Lorimer,
Lyne \& Camilo~(1998)\nocite{llc98}, with a resulting flux density
limit for the pulsar of $S_{600} < 1.1$\,mJy at a search frequency of
$\nu = 600$\,MHz.  This corresponds to $S_{1400} < 0.3$\,mJy for a
median pulsar spectral index of $\alpha = -1.6$ \cite{lylg95}, where
$S_\nu \propto \nu^\alpha$.  At a distance $d = 3.2$\,kpc \cite{rgk+93}
the implied luminosity limit $L_{1400} \equiv S_{1400} d^2 <
3$\,mJy\,kpc$^2$ is somewhat larger than the radio luminosity of at
least one young pulsar \cite{cmg+01}, and the discovery of X-ray
pulsations from \psr\ at the center of 3C58 by Murray et
al.~(2002)\nocite{mss+01} led us to attempt more sensitive searches.

On 2001 October 8 we used the 100\,m Effelsberg telescope to conduct a
search of \psr.  We observed the pulsar position for 12.5\,hr at a
center frequency of 1410\,MHz.  Signals were filtered in
$8\times4$\,MHz-wide channels for each of 2 polarizations, after which
they were detected and digitized every 0.8\,ms.  Samples from
orthogonal polarizations were added in hardware and the 8 total power
time series were written to disk for offline analysis.  We also
recorded short data sets on strong pulsars for calibration purposes.
We analyzed the data in standard fashion (see below for description of
the search at Green Bank), de-dispersing the time series for a number
of trial DMs in the range 0--500\,cm$^{-3}$\,pc, with no significant
candidates found at the expected period.  From these observations we
inferred $S_{1400} < 0.1$\,mJy for a pulsar duty-cycle of $\sim
0.05P$.

On 2002 February 22--23 we used 18\,hr at the Green Bank Telescope to
search for radio emission from \psr.  We divided the time nearly
equally between an observation centered at 1375\,MHz with a Gregorian
focus receiver, followed by one at 820\,MHz with a prime focus
receiver.  The telescope gain and system temperature, including a large
contribution from the SNR, are given in Table~\ref{tab:obs}.  We also
performed short calibration observations of strong pulsars at both
frequencies.  The receivers, sampling 2 orthogonal polarizations, had
band-limiting filters installed of width 150 and 80\,MHz centered on
1375 and 820\,MHz respectively.  The radio frequency (RF) signals were
converted to an intermediate frequency (IF) and transmitted via optical
fibers to the control room, where they were passed to the
Berkeley-Caltech Pulsar Machine (BCPM).

The BCPM is one of several signal processors of similar design
\cite{bdz+97}.  The 2 input IF signals are distributed to an array of
$2\times6$ mixer/filter modules.  These accept signals from 6 local
oscillator modules for baseband mixing and low-pass filtering with the
bandwidth set by subsequent signal processing requirements.  The
signals are sent to $2\times6$ digital filter boards which 4-bit sample
and form detected powers in 16 independent and adjacent frequency
channels.  The BCPM is thus an analog/digital filter bank with
$2\times96$ channels.  Time decimation of the power samples and
optional summing of the 2 IFs (which we used) is accomplished in
combiner boards.  The combiner board output is reduced to 4-bit power
deviation by mean removal and scaling.  The power vector(s) are passed
to the host Sun workstation through an EDT\footnote{Engineering Design
Team, Inc.} DMA card.  The Sun attaches a header and places data on a
disk for offline processing as well as monitoring data
quality\footnote{For further details of BCPM installation at Green Bank
see http://www.gb.nrao.edu/$\sim$dbacker.}.  For our 2 observations we
used the data-taking parameters listed in Table~\ref{tab:obs}.


\section{Data Analysis and Results}\label{sec:results}

Although we knew the pulsar period a priori, we did not know its DM and
chose to use standard search algorithms to analyze the data.  Also,
despite the presence of strong RF interference (RFI) visible in a
spectrum analyzer during data collection (most notably as persistent
radar in the 1400\,MHz band and sporadic but very strong RFI of unknown
origin in the 800\,MHz band) in a first pass we masked no portions of
the fluctuation spectra.  For both observations data from $2^{25}$
decimated time samples (see Table~\ref{tab:obs}) were de-dispersed at
204 trial DMs in the range 0--203\,cm$^{-3}$\,pc (the DM predicted by
the Taylor \& Cordes 1993\nocite{tc93} electron density/distance model
is $\sim 70$\,cm$^{-3}$\,pc).  Each of the 204 resulting time series
was then searched for periodic signals over a range of duty cycles with
an FFT-based code \citep[described in detail by][]{lkm+00} identifying
significant features in the fundamental amplitude spectrum as well as
in spectra with 2, 4, 8, and 16 harmonics folded in.  The 1375\,MHz
data were heavily corrupted by RFI --- in particular the 182nd harmonic
of a radar signal repeating every $\sim 12$\,s coincided with the
expected pulsar rotation frequency.  Therefore we eventually reanalyzed
these data while blanking the radar frequency and its first 150
harmonics in the amplitude spectra.

The 820\,MHz data were clean in the spectral range of interest and a
signal was unambiguously detected at the expected period and clearly
dispersed, peaking at $\mbox{DM} \approx 143$\,cm$^{-3}$\,pc with a
signal-to-noise ratio S/N = 10.9, as shown in Figure~\ref{fig:SNvDM}.
After excision of the radar-related RFI, the 1375\,MHz data yielded a
signal at the same period within the uncertainty peaking at $\mbox{DM}
\approx 140$\,cm$^{-3}$\,pc with S/N = 8.8.


\medskip

\epsfxsize=7.0truecm
\epsfbox{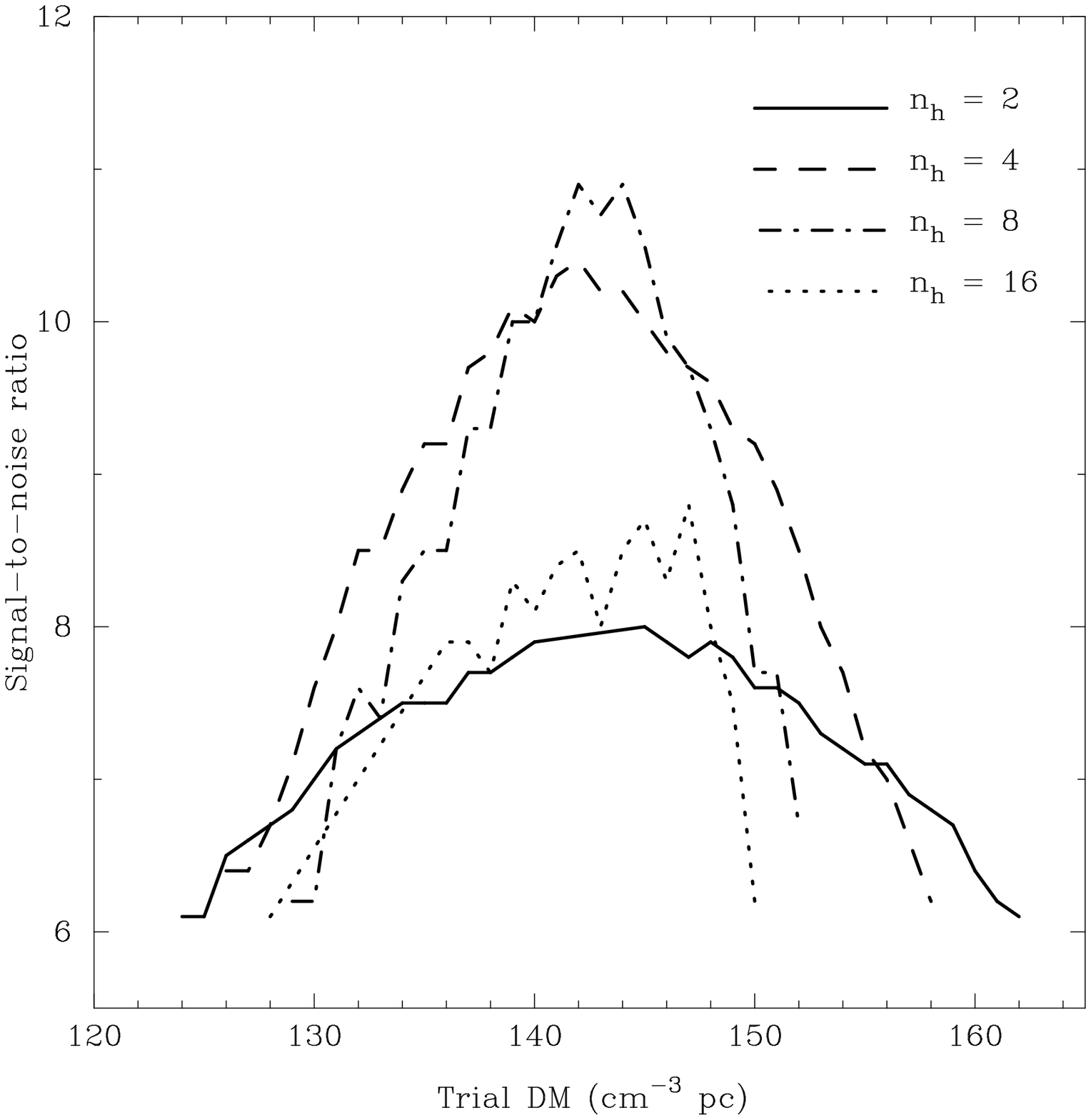}
\figcaption[f1.ps]{\label{fig:SNvDM}
Search signal-to-noise ratio as a function of trial dispersion measure
(DM) for the candidate with period $P \approx 65.69$\,ms in the
820\,MHz data set.  Each curve corresponds to the addition of the
number of harmonics indicated. }

\medskip

We then measured pulse times-of-arrival (TOAs) referenced to the
observatory atomic time standard from both data sets: 6 good TOAs were
obtained in each of the 800 and 1400\,MHz bands with individual
uncertainty $\approx 0.5$\,ms.  A further 17 TOAs of similar quality
were obtained on 8 additional days of observations, with a total data
span of 35\,d.  We used the TOAs and the {\sc tempo} timing
software\footnote{See http://pulsar.princeton.edu/tempo.} to fit for
$P$, $\dot P$, and DM, obtaining the values listed in
Table~\ref{tab:parms}.  The table also lists the celestial coordinates
of \psr\ obtained by Slane et al.~(2002)\nocite{shm02}, which were held
fixed in our timing fit.  Our values of $P$ and $\dot P$ determined
from a phase-connected timing solution spanning 35\,d are much more
precise than, but agree with, the equivalent values obtained by Murray
et al.~(2002)\nocite{mss+01} based on only 2 periods measured 1.2 and
4.4\,yr before: the barycentric period $P$ listed in
Table~\ref{tab:parms} differs by $15 \pm 18$\,ns from that extrapolated
to the epoch of our observations using the $P$ and $\dot P$ determined
by Murray et al., while our $\dot P$ is consistent with theirs at the
$2\,\sigma$ level.  Such a close match of spin parameters over an
interval greater than 4\,yr suggests that no significant period glitch
occurred during this time span.


Both search data sets, de-dispersed at $\mbox{DM} = 141$\,cm$^{-3}$\,pc
and folded at the best search periods, are shown in
Figure~\ref{fig:profs}.  The average pulse profile at both frequencies
consists of a single narrow pulse of width $\approx 0.05 P$ (see
Table~\ref{tab:parms}); such a profile is expected based on
Figure~\ref{fig:SNvDM}, where the search S/N is a maximum for 8
harmonics summed.  There is no evidence for the existence of an
interpulse at either frequency with amplitude $\ga 15\%$ that of the
sharp pulse, in contrast with the X-ray profiles that show an
interpulse with amplitude $\sim 30\%$ that of the main pulse
\cite{mss+01}.


\medskip

\epsfxsize=8.0truecm
\epsfbox{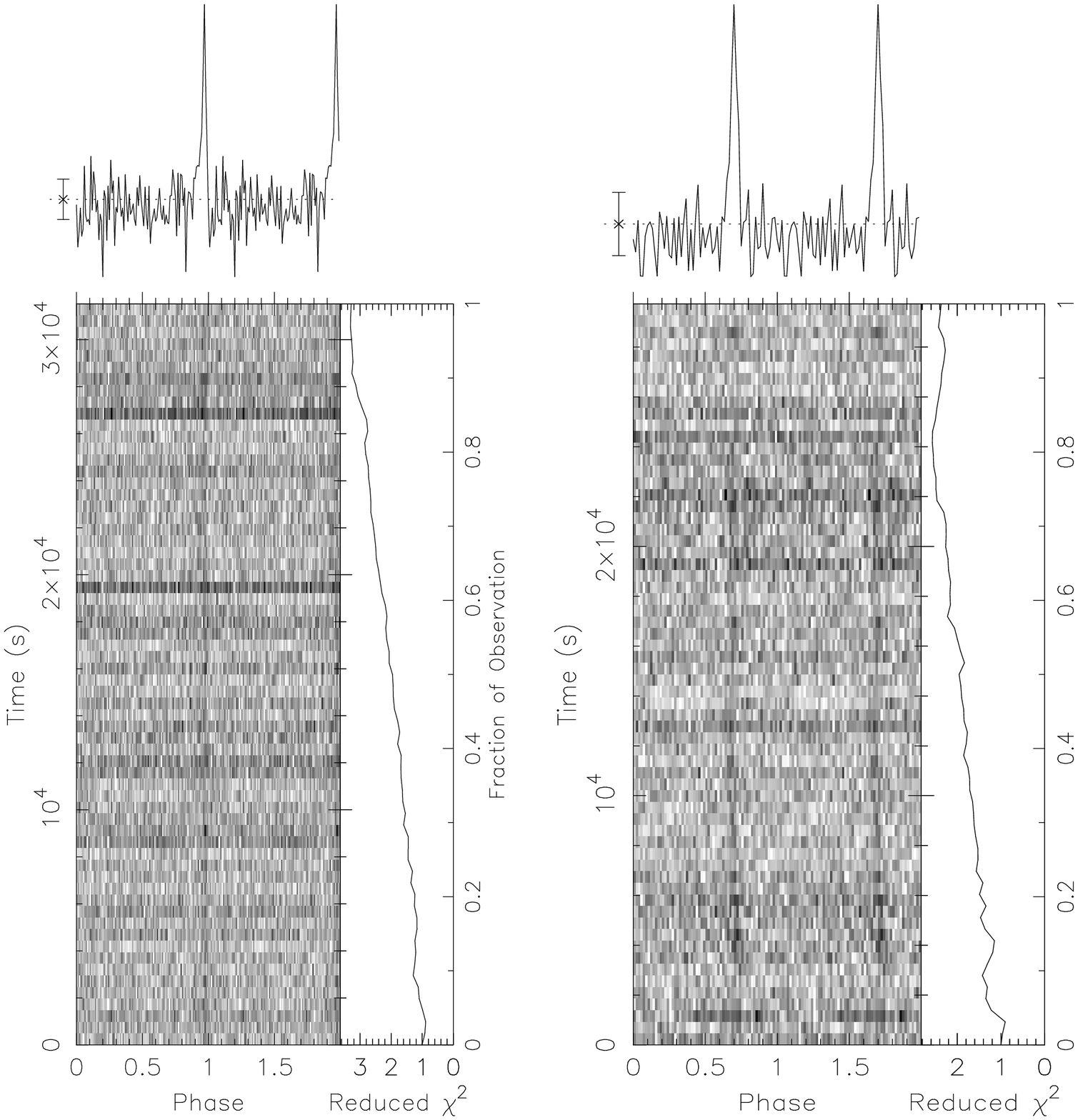}
\figcaption[f2.eps]{\label{fig:profs}
Pulse profiles of \psr\ displayed as a function of time ({\em
bottom\/}) and summed ({\em top\/}).  {\em Left\/}: Data at a center
frequency of $\nu = 1375$\,MHz presented with 100 phase bins.  The
pulse full-width at half maximum (FWHM) is $2.3 \pm 0.3$\,ms.  {\em
Right\/}:  Data at $\nu = 820$\,MHz displayed with 60 bins.  The pulse
$\mbox{FWHM} = 3.8 \pm 0.4$\,ms.  All data were de-dispersed at
$\mbox{DM} = 141$\,cm$^{-3}$\,pc and one phase bin in each panel
matches the dispersion smearing resolution of the respective profile.
Two full periods are shown at each frequency for clarity.  The
increasing/decreasing slope of each reduced $\chi^2$ curve traces
pulsation significance as measured against a constant flux (or
non-pulsed) model.  Decreases in reduced $\chi^2$ are most likely
caused by interstellar scintillation or RFI. }

\medskip

Note that the S/N in the 820\,MHz data appears to be smaller near the
beginning of the integration and again towards the end
(Fig.~\ref{fig:profs}), possibly as a consequence of scintillation.
Assuming that the search S/Ns in the 2 long integrations reflect the
respective long-term time-averaged flux densities in both frequency
bands, we have integrated the area under the pulse profiles and
measured their baseline rms to determine the respective flux
densities.  We converted to a Jy scale using the observing parameters
listed in Table~\ref{tab:obs} together with the radiometer equation
\cite{dic46}.  In this manner we estimate that $S_{1400} \sim
45\,\mu$Jy and $S_{800} \sim 130\,\mu$Jy, with $\sim 25\%$ accuracy.
The resulting spectral index in the 800--1400\,MHz range is $\alpha
\sim -2.1 \pm 0.6$, and the radio luminosity is $L_{1400} \sim
0.5\,(d/3.2\,\mbox{kpc})^2\,$mJy\,kpc$^2$.  These values will be much
improved by the averaging of several calibrated observations.

The 947\,yr-old Crab pulsar was detected initially through its
individual ``giant'' pulses rather than through its time-averaged
pulsed emission \cite{sr68}.  As the emission of giant pulses may be
related to the magnetic field strength at the pulsar light cylinder,
$B_{lc}$ \citep[e.g.,][]{cstt96}, and \psr\ has the 10th largest value
of $B_{lc}$ among known pulsars, we searched for aperiodic, dispersed
pulses from \psr.  No excess of single pulses was found at the DM of
the pulsar.  Inferring a quantitative limit is complicated by
significant levels of RFI, and this search will be reported in detail
elsewhere.

\section{Discussion}\label{sec:disc}

We have discovered radio pulsations from the direction of \psr\ in
SNR~3C58 in an unbiased search of 2 independent data sets acquired at
center frequencies of 820 and 1375\,MHz.  The pulsations are clearly
dispersed and have a precisely determined period matching that observed
from \psr\ at X-ray wavelengths \cite{mss+01}.  We have therefore
discovered radio pulsations that are unambiguously from \psr\ which
supplants the Crab pulsar as the youngest neutron star with known radio
emission.

The $\mbox{DM} = 141$\,cm$^{-3}$\,pc of \psr, located at $(l,b) =
(130\fdg72, 3\fdg08)$, is approximately twice that predicted by the
Taylor \& Cordes (1993)\nocite{tc93} model for the usually accepted
distance to 3C58 (3.2\,kpc, determined from H\,{\sc i} absorption
measurements\footnote{Depending on data set and Galactic rotation curve
used, the inferred distance ranges from 2.6\,kpc to ``just beyond''
3.2\,kpc \citep[see][and references therein]{rgk+93}.  We adopt $d =
3.2$\,kpc in this Letter.}).  This implies an average free electron
density to 3C58 of $n_e \approx 0.044$\,cm$^{-3}$.  The neutral column
density to the SNR is $3\times10^{21}$\,cm$^{-2}$ (Helfand, Becker, \&
White 1995\nocite{hbw95}), implying an ionized fraction of roughly
15\%, which is somewhat higher than might be expected towards the
Galactic plane.  However, the SNR lies on the eastern edge of the large
W3/W4/W5 complex of H\,{\sc ii} regions (e.g., Reynolds, Sterling, \&
Haffner 2001\nocite{rsh01}), and in fact may be on the edge of a cavity
being ionized by the star-forming regions (R.~J. Reynolds and G.
Madsen, private communication).  It is not known if the progenitor of
SN~1181 belonged to, or was a runaway star from, this region of star
formation.  In any case, the relative locations of 3C58 and W3/W4/W5
may account for the apparent excess of ionized material towards the
pulsar.

We note that the reported detection of \psr\ with $\mbox{DM} =
24$\,cm$^{-3}$\,pc at a frequency of 110\,MHz (Malofeev, Malov, \&
Glushak 2001\nocite{mmg01}) is incompatible with the results we
present:  the dispersion smearing timescale for a signal with
$\mbox{DM} = 141$\,cm$^{-3}$\,pc within one of the 20\,kHz-wide
spectral channels used by Malofeev et al.~(2001)\nocite{mmg01} at
110\,MHz is 17\,ms.  De-dispersing and summing their 32 individual such
channels with $\mbox{DM} = 24$\,cm$^{-3}$\,pc would result in an
overall smearing of the pulse far in excess of $P = 65$\,ms, rendering
pulsations unobservable from \psr.

The main X-ray pulse from \psr\ has essentially the same sharp
structure and width \citep[$\approx 0.04P$;][]{mss+01} as the radio
pulse (Fig.~\ref{fig:profs} and Table~\ref{tab:parms}).  This
morphology suggests that the X-ray emission is likely magnetospheric in
nature, but absolute alignment of radio and X-ray profiles will provide
further clues for understanding the emission mechanism(s).  Considering
the obvious existence of magnetospheric emission, we might expect this
pulsar to be a $\gamma$-ray source.  The flux sensitivity of the EGRET
instrument to emission from \psr, using a photon index $\Gamma = 2$
typical of $\gamma$-ray-emitting pulsars, is $F_\gamma \sim
5\times10^{32}$\,erg\,s$^{-1}$\,kpc$^{-2}$
\citep[$>100$\,MeV;][]{hbb+99}, corresponding to a luminosity limit
$L_\gamma \equiv F_\gamma d^2 \la 5\times10^{33}$\,erg\,s$^{-1}$.  With
spin-down luminosity $\dot E = 2.7\times10^{37}$\,erg\,s$^{-1}$, its
efficiency at converting rotational power into beamed high-energy
$\gamma$-rays is $\eta \equiv L_\gamma/\dot E \la 2\times10^{-4}$.
This compares to $\eta \sim 10^{-4}$ for the Crab pulsar, with the
largest value of $\dot E$ known in the Galaxy, and larger efficiencies
for (generally older) pulsars with smaller values of $\dot E$ such as
Vela ($\eta \sim 6\times10^{-4}$) and Geminga ($\eta \sim 0.02$).  If
\psr\ follows the approximate $\eta \propto {\dot E}^{-1/2}$ relation
that seems to apply to known $\gamma$-ray-emitting pulsars
\cite{tbb+99}, its flux may be just below the detection threshold of
EGRET and should be detected by next-generation observatories such as
{\em GLAST}.

The discovery of extremely weak radio pulsations from \psr\ highlights
the vexing issue of the true luminosity distribution of (particularly
young) pulsars.  The past year has witnessed the discovery of 3 very
young ($\tau_c \la 10$\,kyr) radio-emitting pulsars with $0.5 \la
L_{1400} \la 3$\,mJy\,kpc$^2$, 1--2 orders of magnitude less luminous
than any young pulsars previously known \citep[][this
Letter]{hcg+01,cmg+01}.  It now seems abundantly clear that we have yet
to probe the depths of the luminosity function of young pulsars.  The
state-of-the-art for finding young radio pulsars ($\sim 10$\,hr
integrations at telescopes with gain $\ga 1$\,K\,Jy$^{-1}$ and
bandwidths 100--300\,MHz at $\nu \sim 1400$\,MHz) yields limits usually
no better than $L_{1400} \sim 1$\,mJy\,kpc$^2$.  Nevertheless, many
worthy targets remain to be searched at this level.  Improving the
existing limits helps to constrain better the ``beaming fraction'' for
young pulsars \citep[e.g.,][based in part on the non-detection of the
pulsar in 3C58 with a limit of 0.15\,mJy at 1400\,MHz, estimated a
beaming fraction of 0.6 which needs to be revised in light of recent
discoveries]{fm93}, and is important for building an accurate census of
Galactic neutron stars.  In this regard, the detection of \psr, an
820\,yr-old pulsar, at a frequency of 820\,MHz --- only the 2nd pulsar
ever discovered in the 800\,MHz band \cite{dmd+88} --- suggests that
the availability of the magnificent GBT provides a new region of phase
space with significant, and as yet untapped, discovery potential.

\acknowledgments

We are deeply grateful to the dedicated staff of NRAO, and in
particular to the GBT project team, whose hard work has made the Robert
C. Byrd Green Bank Telescope a great research instrument.  We thank
Bryan Jacoby and Stuart Anderson for assistance with the BCPM, Jay
Lockman for helpful discussions, and David Helfand for early
communication of discovery of X-ray pulsations.  The National Radio
Astronomy Observatory is a facility of the National Science Foundation
operated under cooperative agreement by Associated Universities, Inc.
FC acknowledges support from NASA grants NAG~5-9095 and NAG~5-9950.
IHS is a Jansky Fellow.  DRL is a University Research Fellow funded by
the Royal Society.  SMR is a McGill University Tomlinson Fellow.



\clearpage

\begin{deluxetable}{lll}
\tablecaption{\label{tab:obs}Summary of Observations and Analyses at GBT }
\tablecolumns{3}
\tablewidth{0pc}
\tablehead{
\colhead{Observing frequency (MHz)} &
\colhead{1375}                      &
\colhead{820}                       \\}
\startdata
Telescope gain (K\,Jy$^{-1}$)\dotfill             & 2.0  & 2.0   \\
System temperature on cold sky (K)\dotfill        & 20   & 25    \\
Galactic synchrotron temperature (K)\dotfill      & 2    & 9     \\
SNR~3C58 temperature (K)\dotfill                  & 64   & 67    \\
Spectral channel width (MHz)\dotfill              & 1.4  & 0.5   \\
Sample time ($\mu$s)\dotfill                      & 50   & 72    \\
Integration time (hr)\dotfill                     & 8.75 & 8.25  \\
Effective time resolution of search (ms)\dotfill  & 0.9  & 0.864 \\
Effective integration time of search (hr)\dotfill & 8.4  & 8.1   \\
\enddata
\end{deluxetable}


%
%
\begin{deluxetable}{ll}
\tablecaption{\label{tab:parms}Parameters of \psr\ }
\tablecolumns{2}
\tablewidth{0pc}
\tablehead{
\colhead{Parameter}   &
\colhead{Value}     \\}
\startdata
R.A. (J2000)\tablenotemark{a}\dotfill          & $02^{\rm h}05^{\rm m}37\fs92$\\
Decl. (J2000)\tablenotemark{a}\dotfill         & $+64\arcdeg49'42\farcs8$     \\
Period, $P$ (ms)\dotfill                       & 65.68638162(2)               \\
Period derivative, $\dot P$\dotfill            & $1.9393(4)\times10^{-13}$    \\
Epoch (MJD [TDB])\dotfill                      & 52345.0                      \\
Dispersion measure, DM (cm$^{-3}$\,pc)\dotfill & $140.7(3)$                   \\
Pulse FWHM at 1375\,MHz (ms)\dotfill           & $2.3 \pm 0.3$                \\
Pulse FWHM at 820\,MHz (ms)\dotfill            & $3.8 \pm 0.4$                \\
Flux density at 1375\,MHz, $S_{1400}$ ($\mu$Jy)\dotfill & $\sim  45$          \\
Flux density at 820\,MHz, $S_{800}$ ($\mu$Jy)\dotfill   & $\sim 130$          \\
Distance of SNR~3C58, $d$ (kpc)\dotfill        & $\approx 3.2$                \\
Derived parameters:                                      &                    \\
~~Radio luminosity, $L_{1400}$ (mJy\,kpc$^2$)\dotfill    & $\sim 0.5$         \\
~~Spectral index, $\alpha$\dotfill                       & $\sim -2.1 \pm 0.6$\\
~~Mean free electron density, $n_e$ (cm$^{-3}$)\dotfill  & $\approx 0.044$    \\
\enddata
\tablecomments{Numbers in parentheses represent $1\,\sigma$
uncertainties in the least-significant digits of fitted parameters.}
\tablenotetext{a}{Position known with $\approx 0\farcs5$ accuracy from
\chandra\ data (Slane et al. 2002)\nocite{shm02}.}
\end{deluxetable}





\end{document}